\documentclass[a4paper,11pt]{article}
\pdfoutput=1 

\usepackage{jinstpub} 
\usepackage{siunitx}

\title{\boldmath A 50 ps resolution monolithic active pixel sensor without internal gain in SiGe BiCMOS technology}

\author[a,1]{G. Iacobucci,\note{Corresponding author.}}
\author[b]{R. Cardarelli,}
\author[a]{S. D\'ebieux,}
\author[a]{F.A. Di Bello,}
\author[a]{Y. Favre,}
\author[a]{D. Hayakawa,}
\author[c]{M. Kaynak,}
\author[a, d]{M. Nessi,}
\author[a]{L. Paolozzi,}
\author[c]{H. R\"ucker,}
\author[a]{DMS Sultan,}
\author[a]{and P. Valerio}

\affiliation[a]{D\'epartement de Physique Nucl\'eaire et Corpusculaire, Universit\'e de Gen\`eve,\\ Quai Ernest-Ansermet, Geneva, Switzerland}
\affiliation[b]{INFN Section of Roma Tor Vergata,\\ Via della ricerca scientifica 1, Roma, Italy}
\affiliation[c]{IHP - Leibniz-Institut f\"ur innovative Mikroelektronik \\ Im Technologiepark 25, Frankfurt (Oder), Germany}
\affiliation[d]{CERN,\\
CH-1211 Geneve 23, Switzerland}

\emailAdd{giuseppe.iacobucci@unige.ch}

\abstract{A monolithic pixelated silicon detector designed for high time resolution  has been produced  in the SG13G2 130 nm SiGe BiCMOS technology of IHP.  This proof-of-concept chip contains hexagonal pixels of \SI{65}{\micro\meter} and \SI{130}{\micro\meter} side.
The SiGe front-end electronics implemented
provides an equivalent noise charge of 90  and 160 $ \mathrm{e^{-}} $ for a pixel capacitance of \SI{70} and \SI{220}{\femto\farad}, respectively, and a total time walk of less than \SI{1}{\nano\second}.
 Lab measurements with a $ \mathrm{^{90}Sr} $ source
 show a time resolution of the order of \SI{50}{\pico\second}. 
This result  is competitive with silicon technologies that integrate an avalanche gain mechanism. 
}


\begin{document}
\maketitle
\flushbottom

\section{Fast, low noise SiGe electronics and sensor design for sub-100ps timing}
\label{sec:intro}

\subsection{Introduction}

Our previous researches \cite{TTPET_discrete}\cite{TTPET_first-chip}\cite{pierpaolo_chip}\cite{TTPET_demonstrator} proved the feasibility of a fully-efficient monolithic pixel detector in SiGe BiCMOS technology with 100 ps RMS time resolution for minimum ionising particles (MIPs).
The result of \cite{TTPET_demonstrator} was obtained with a simple design that integrated the  pixel matrix inside the guard-ring, while keeping the electronics in the periphery of the chip and referring the substrate to ground.
The  $ \mathrm{500\times500} $ \SI{}{\micro\meter\squared} area of the pixels of that chip, designed for the TT-PET project  \cite{TTPET_project}, was sufficient to utilise it  without impacting the intrinsic resolution of positron-emission tomography. That relatively-large pixel area resulted in a pixel capacitance of 750 fF that  was limiting the  noise performance of the front-end.
Since  Cadence Spectre simulations showed that the Equivalent Noise Charge (ENC) of the preamplifier implemented in that chip decreases linearly with the pixel capacitance down to \SI{75}{\femto\farad} \cite{pierpaolo_chip}, we investigated the possibility for our SiGe  BiCMOS frontend to achieve time resolutions below \SI{100}{\pico\second}, 
even in absence of an internal gain mechanism, simply by reducing the pixel capacitance.


\subsection{Pixel matrix design}

The reduction of the pixel capacitance requires to decrease  the pixel size and  reduce the signal routing distance to the preamplifier. 
In the prototype  described here, realized in the SG13G2 130nm BiCMOS technology~\cite{SG13G2} of IHP, this is achieved by placing the electronics in triple wells inside the guard rings.  

\begin{figure}[htbp]
\centering
\includegraphics[width=.75\textwidth,trim=0 0 0 0]{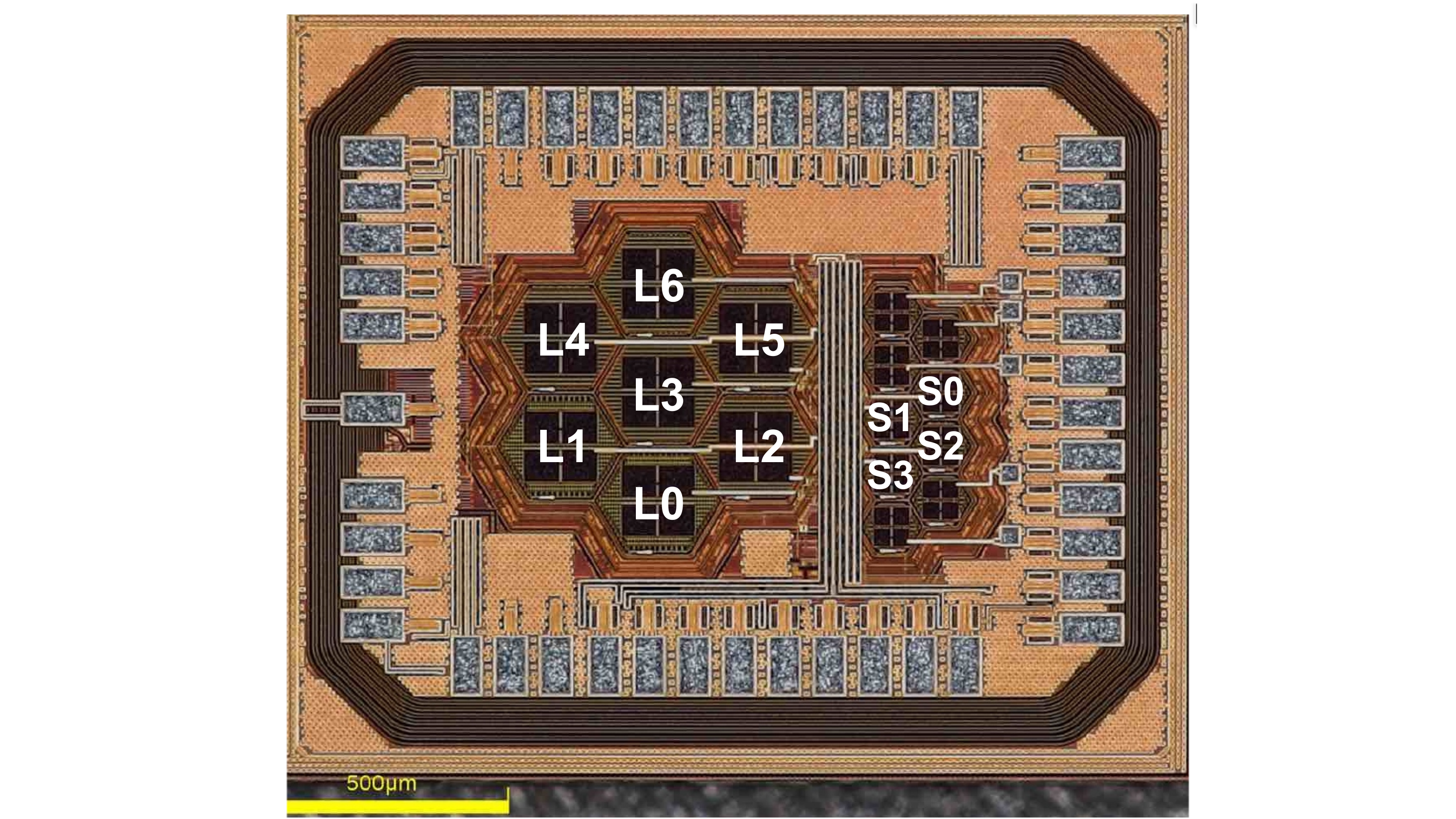}
\caption{\label{fig:chip} 
Photograph of the monolithic chip prototype in SG13G2 IHP technology containing two matrices of hexagonal pixels. 
The larger-surface matrix is made of seven pixel of \SI{130}{\micro\meter} side (pixels L0-L6) that share four outputs channels; pixel L3 and L5, in particular, are connected in OR to the same output driver. 
The second matrix, visible on the right, has nine \SI{65}{\micro\meter}-side pixels, four of which (S0-S3) are connected to the front-end.
The front-end electronics is placed in the region between the two matrices.}
\end{figure}

Figure \ref{fig:chip} shows the prototype chip. It contains two matrices of hexagonal pixels with side of either \SI{130} or \SI{65}{\micro\meter} and areas corresponding to square pixels of approximately \SI{210} and \SI{105} {\micro\meter}, respectively. The hexagonal shape of the pixels was chosen to increase the angles at the corners of the pixels, therefore reducing the electric field at the nearby surface  and limit the risk of an early break-down due to the P-stop layer in the low-resistivity substrate. 
TCAD\footnote{https://www.synopsys.com/silicon/tcad.html} simulations were used to estimate the pixel capacitance, which resulted to be  approximately \SI{220}{\femto\farad} for the large pixel and \SI{70}{\femto\farad} for the small ones, for a high voltage of \SI{140}{\volt} that provides a depletion depth of \SI{26}{\micro\meter}.

\begin{figure}[htbp]
\centering
\includegraphics[width=.18\textwidth,trim=0 0 0 0, angle=90]{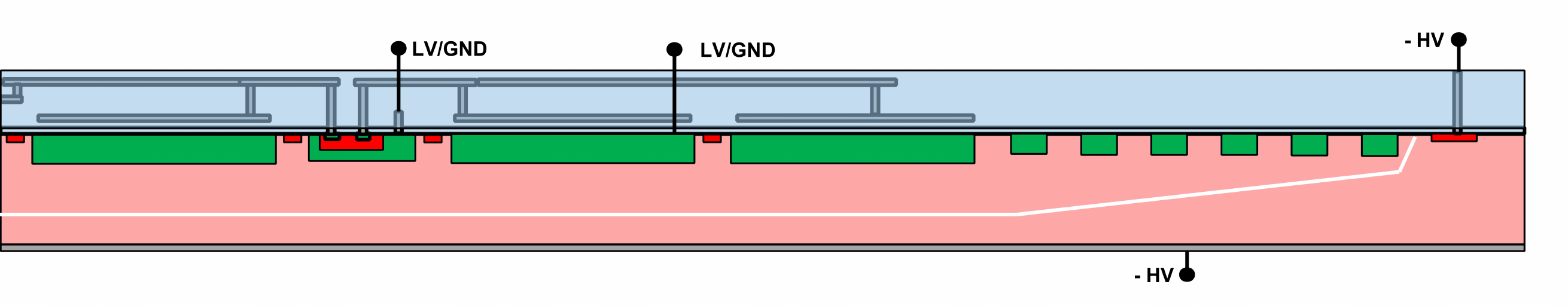}

\vspace{10pt}
\includegraphics[width=.7\textwidth,trim=0 0 0 0, angle=0]{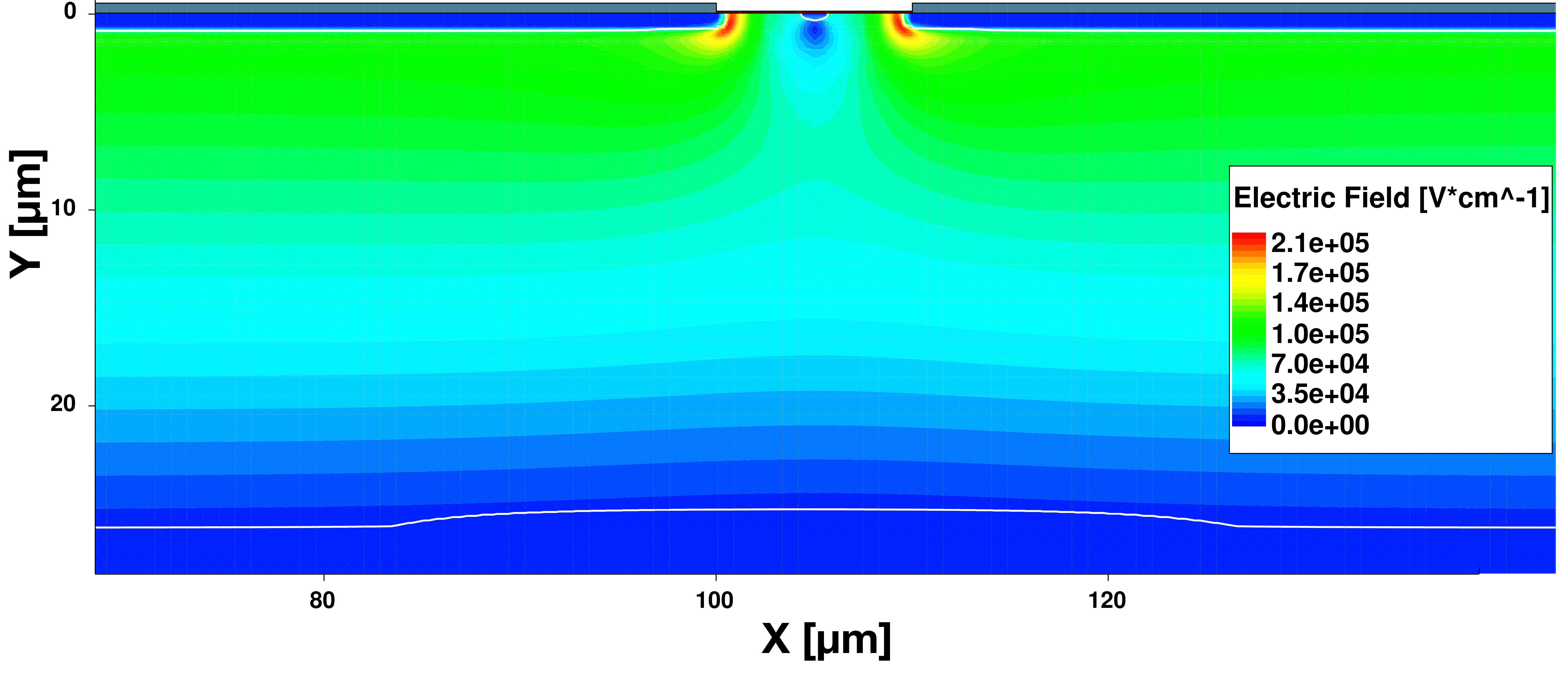}
\caption{\label{fig:cross_section} 
(Top) Conceptual cross section of the monolithic pixel prototype. The P-doped regions are shown in red, while  the N-doped regions in green. The white line gives an indication of the limit of the depletion region. 
The electronics inside the triple wells and the pixel N-well are referred to the positive low-voltage.
(Bottom) TCAD simulation of the Electric field in the inter-pixel region of the sensor. The white line shows the boundary of the depletion region.
The high electric field at the edge of the pixel will be reduced in future prototypes by the use of a higher-resistivity substrate.
}
\end{figure}

The top panel of Figure \ref{fig:cross_section} shows a conceptual cross section of the detector.
The P-doped substrate, in the technology standard resitivity of \SI{50}{\ohm\centi\meter}, is connected to the negative high voltage from the top surface outside the guard-rings and from the chip backside, 
while the pixels and the triple wells containing the electronics are referred to the positive low-voltage. During operation, the HV connection on the chip backside is AC coupled to ground.
The chip was thinned to \SI{60}{\micro\meter} to reduce the resistance in series to the preamplifier input circuit. 
Currents below \SI{10}{\nano\ampere\per\centi\meter\squared} were measured when the chip was operated at bias voltages up to \SI{140}{\volt} to take the data presented here.

Probe-station tests showed that for HV $\approx$ 160 V or larger a current rising up to \SI{700}{\nano\ampere} appears after two or more days of continuous operation. This current disappears if the detector is powered off for several hours. After investigations, it was realised that  this effect could be due to the very high electric field inside the shallow-trench insulation at the chip surface, which is caused by the necessity to deplete a low-resistivity silicon wafer \footnote{This current can be eliminated by using  higher-resistivity wafers that would produce a more uniform electric field.
}.
This current is too low to affect the noise of the pixel detector and  impact  its timing performance. 

The depth of the depletion layer  is not controlled by a backside P+ implantation, but it is limited by the low  resistivity of the wafer. 
This solution greatly simplified the production of this proof-of-concept detector, 
although it slightly degrades the uniformity of the timing performance  
due to the lower electric field at the boundary between two pixels.
This effect is illustrated by the TCAD simulation shown in the bottom panel of Figure \ref{fig:cross_section}.

\subsection{SiGe HBT based front-end}

\begin{figure}[htbp]
\centering
\includegraphics[width=.47\textwidth,trim=0 0 30 25, clip]{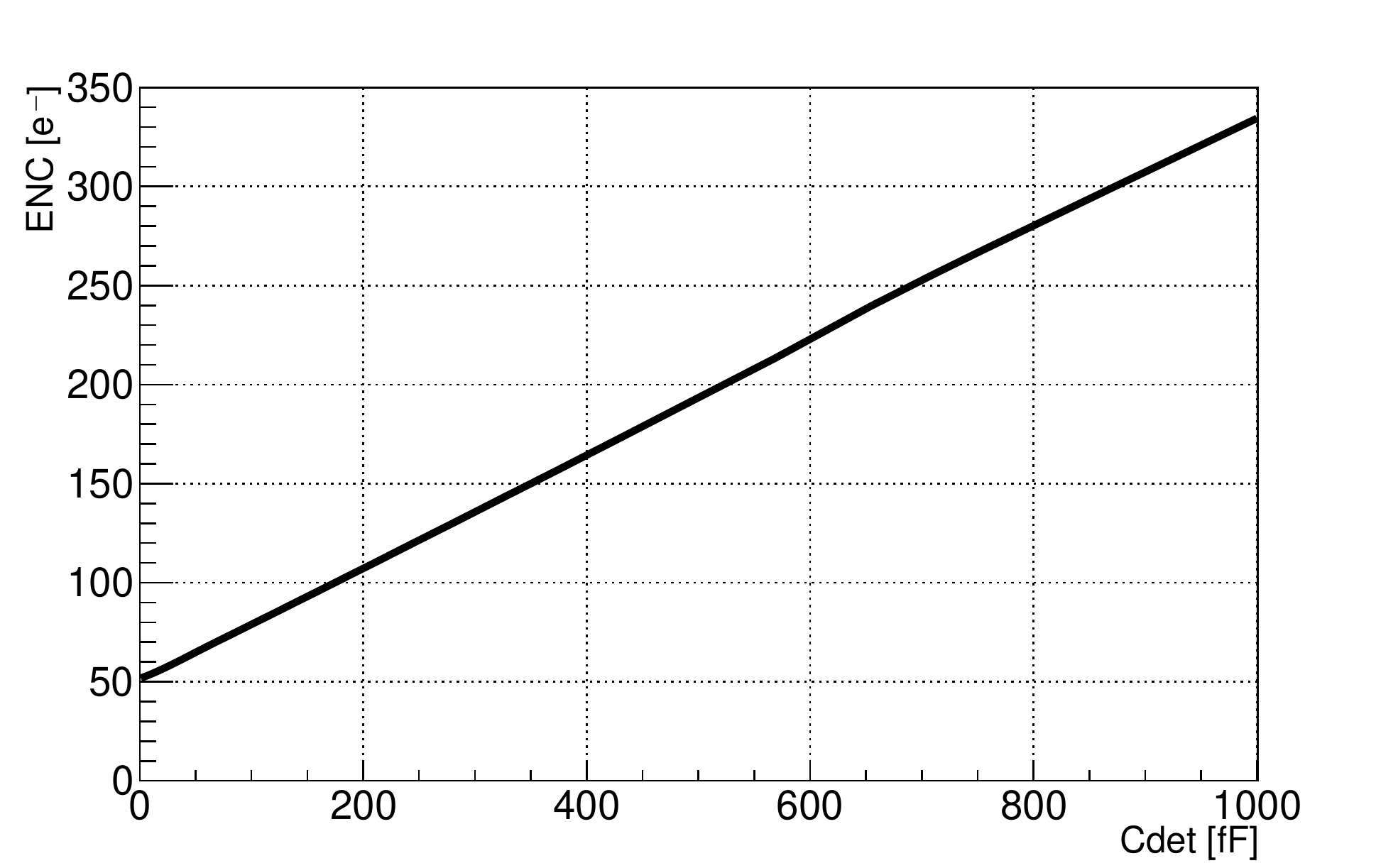}
\qquad
\includegraphics[width=.47\textwidth,trim=0 0 30 25, clip]{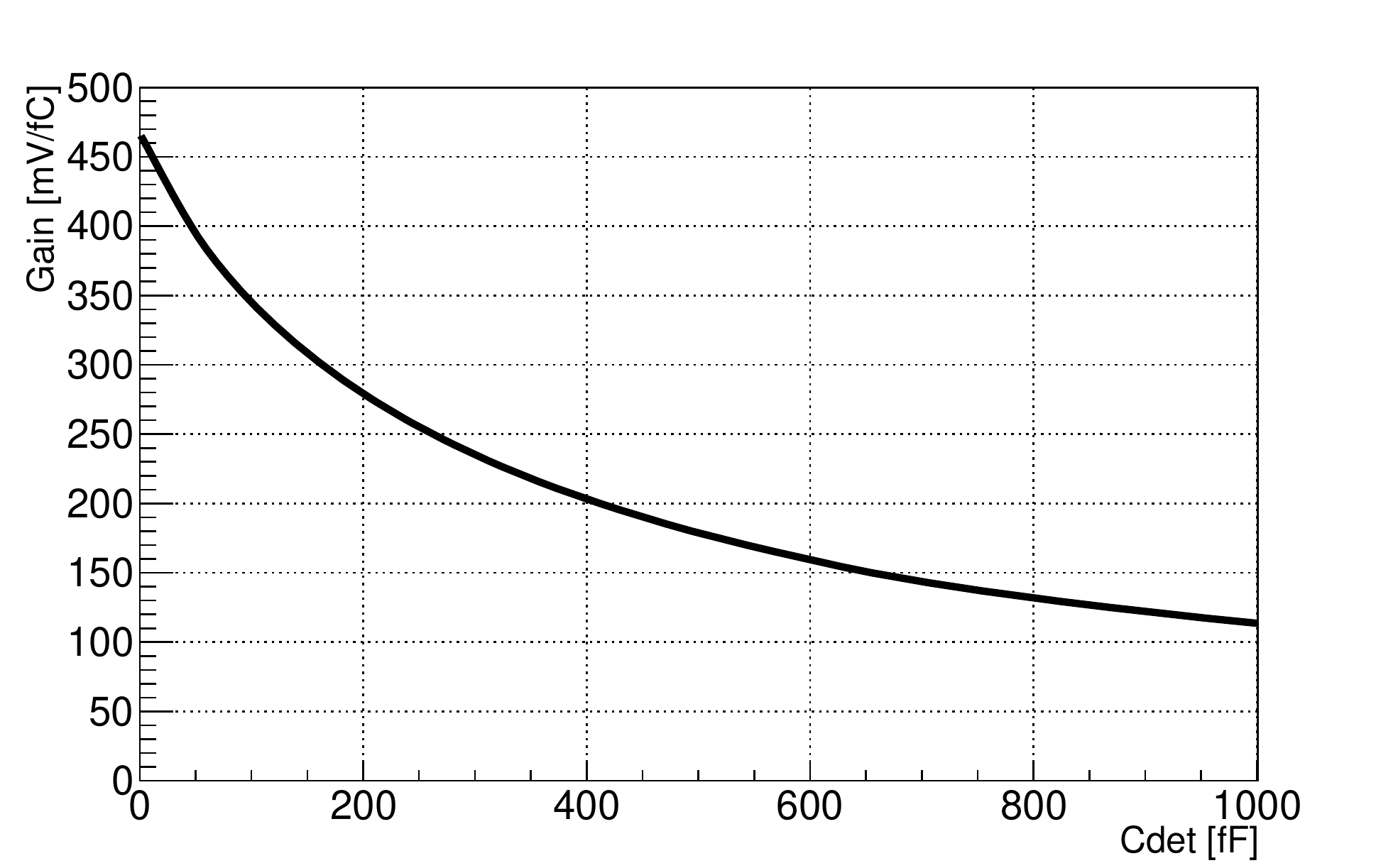}
\caption{\label{fig:ENC_gain}Cadence Spectre simulation of the preamplifier realised in SG13G2 IHP technology.
(Left)   ENC vs.\!\! detector capacitance;  
(Right) charge gain vs.\!\! detector capacitance. 
The capacitance obtained from the TCAD simulation of the large-size  pixels is approximately \SI{220}{\femto\farad} and that of the small pixel \SI{70}{\femto\farad}.
}
\end{figure}

The preamplifier design is based on the one developed for the TT-PET project \cite{pierpaolo_chip}, with minor circuital adaptations to increase the gain and  allow for the placement inside a deep N-well. 
Figure \ref{fig:ENC_gain} shows the ENC and gain of the amplifier as predicted by Cadence Spectre simulation.

The amplifier is DC connected to a CMOS-based open-loop 3-stage discriminator. The signal from the discriminator is sent directly to the output via a low-voltage differential driver. 
Both the Time Of Arrival (TOA) and the Time Over Threshold (TOT) of the amplifier output can be measured from the discriminated signal. Due to the limited number of I/Os available on the chip,  the pixels share four output channels, 
as described in Figure \ref{fig:chip}. 
In order to be able to turn on/off single pixels, each of them has a dedicated discrimination threshold.

\section{Experimental setup and calibrations}
\label{sec:setup}

A $ \mathrm{^{55}Fe} $ radioactive source, that generates \SI{5.9}{\kilo\electronvolt} photons, and a $ \mathrm{^{109}Cd} $, that generates  photons of 22 keV and 25 keV relevant for this study, were used to measure the performance of the preamplifier. 

\begin{figure}[htbp]
\centering 
\qquad
\includegraphics[width=.70\textwidth,trim= 0 0 0 0, clip]{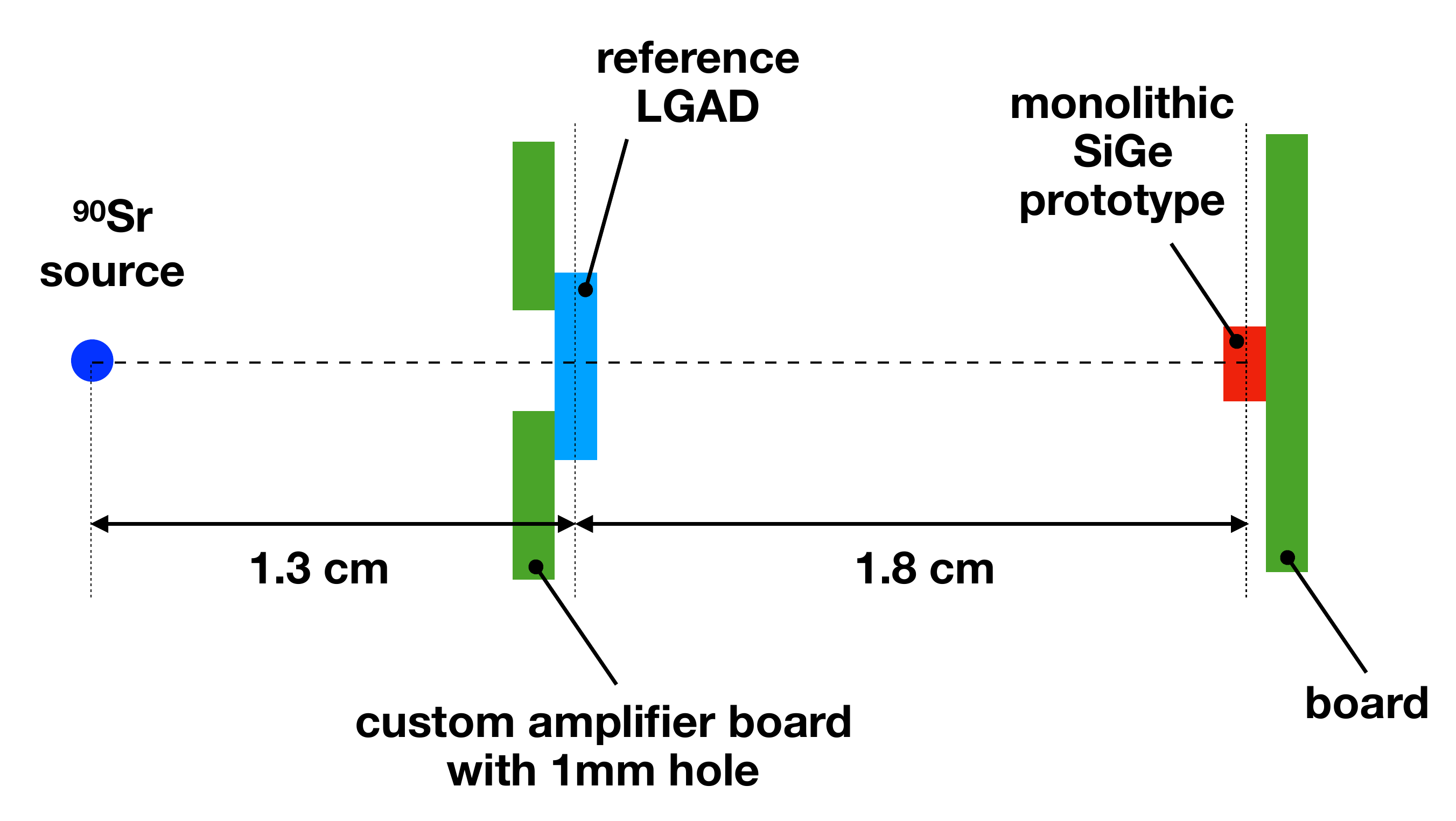}
\caption{\label{fig:setup} Test setup used for the TOF measurements showing the $ \mathrm{^{90}Sr} $ source, the reference LGAD glued on a purposely-designed amplifier board with a 1 mm hole to allow particles to pass through and the monolithic prototype chip under test.}
\end{figure}

The measurement of the time resolution required the use of an LGAD detector~\cite{LGAD_FBK} to provide a reference time. 
The LGAD was glued on a purposely-designed amplifier board, with a 1mm-wide opening under the LGAD to let the electrons from a $ \mathrm{^{90}Sr} $ source reach our chip and enable a Time-Of-Flight (TOF) measurement (see Figure~\ref{fig:setup}). 
The reference LGAD used for this test has a time resolution of \SI{50}{\pico\second} RMS~\cite{LGAD_FBK}.
To verify that the time resolution of the reference LGAD is not affected by the custom amplifier board, the TOF between two LGADs was measured using the same setup.
Our measurement confirmed the time resolution found in the literature.

\section{Results}
\label{sec:results}

\subsection{Performance of the front-end electronics} 

\begin{figure}[htbp]
\centering 
\includegraphics[width=.47\textwidth,trim=0 0 30 20, clip]{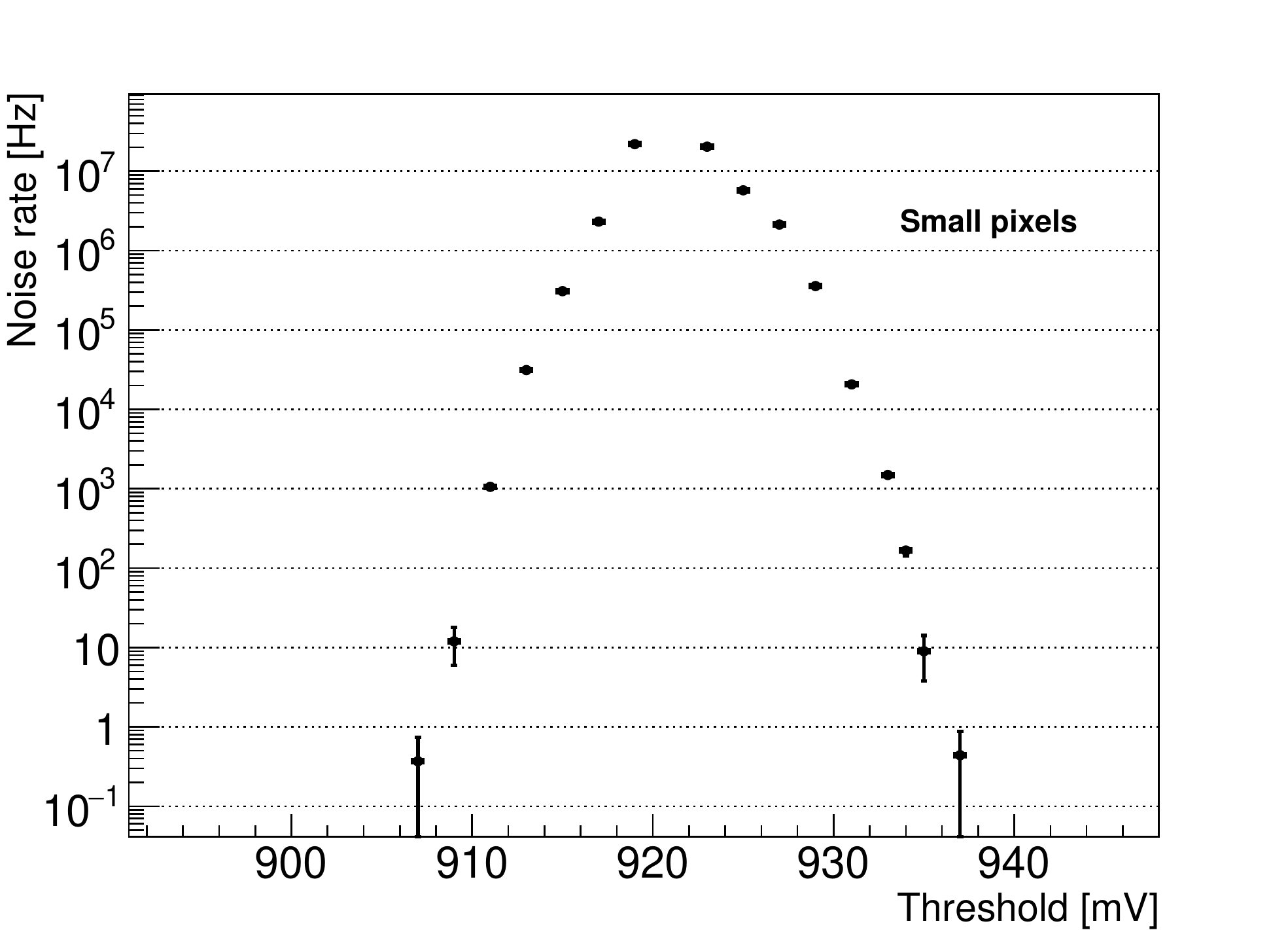}
\qquad
\includegraphics[width=.47\textwidth,trim=0 0 30 20, clip]{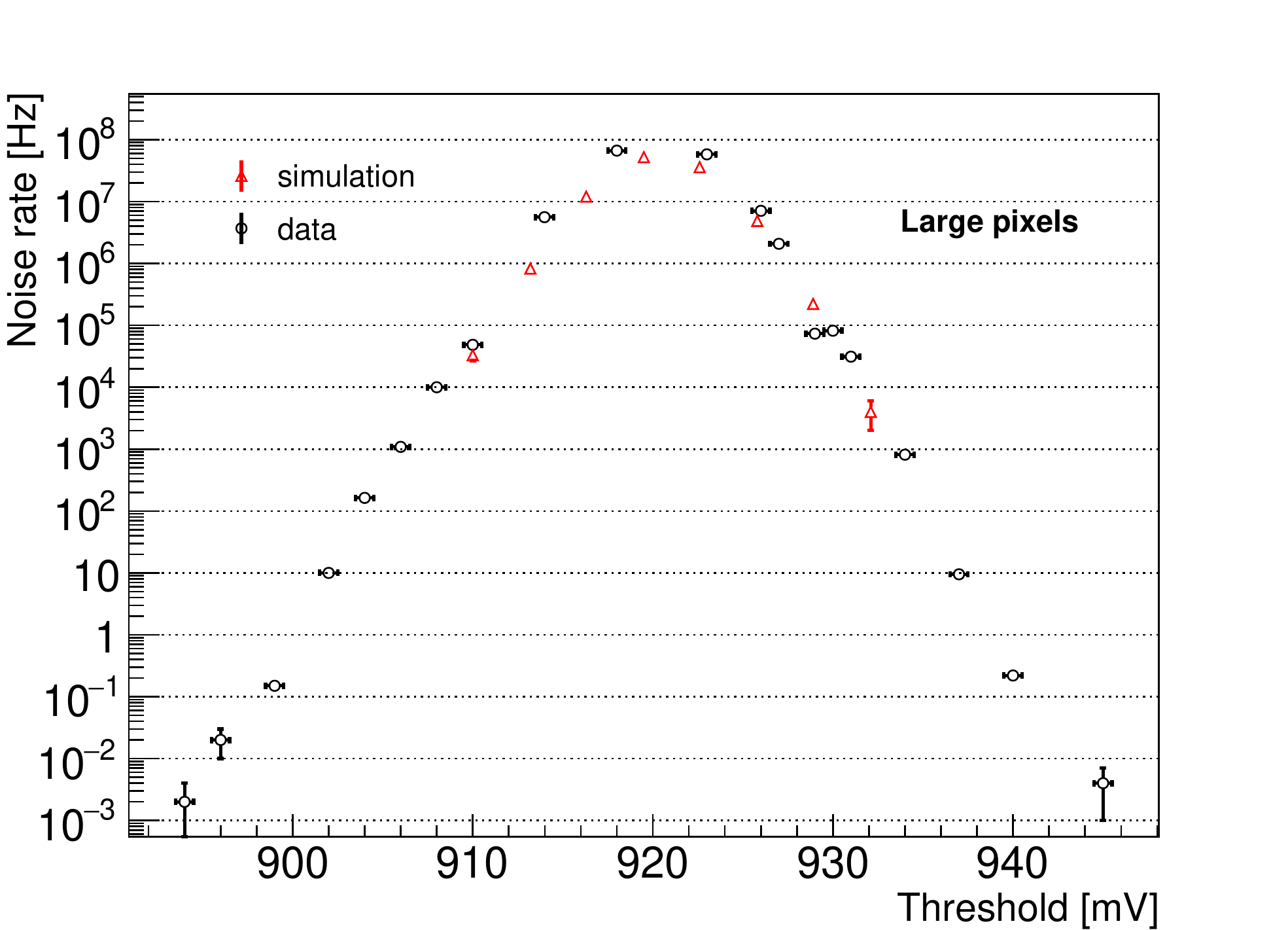}
\caption{\label{fig:noise} 
Noise hit rate as a function of the discrimination threshold measured  for (left) the small pixel S0 and (right) the large pixel L3. For thresholds below \SI{920}{\milli\volt} the discrimination on the oscilloscope was done for a negative pulse slope.
The red points represent the expected noise rate obtained with Cadence Spectre simulations.}
\end{figure}

The left panel of Figure \ref{fig:noise} shows the noise hit rate of the amplifier as a function of the discriminator threshold measured in the case of the small pixel S0 of Figure \ref{fig:chip}, both for a positive (above the baseline) and a negative threshold. 
The right panel of the same figure shows the noise hit rate for the large pixel L3.
The noise hit rate measurements give an RMS voltage noise of \SI{2.67}{\milli\volt} and \SI{2.99}{\milli\volt} for the small and the large pixels, respectively.
However, since the discriminator hysteresis acts as a filter,
the actual RMS amplifier noise was obtained from Cadence Spectre simulations, and it is found to be $ \sigma_V =$ \SI{4.0}{\milli\volt} for the small pixels and $ \sigma_V =$ \SI{4.7}{\milli\volt} for the large ones.

\begin{figure}[htbp]
\centering 
\includegraphics[width=.49\textwidth,trim=0 0 0 0, clip]{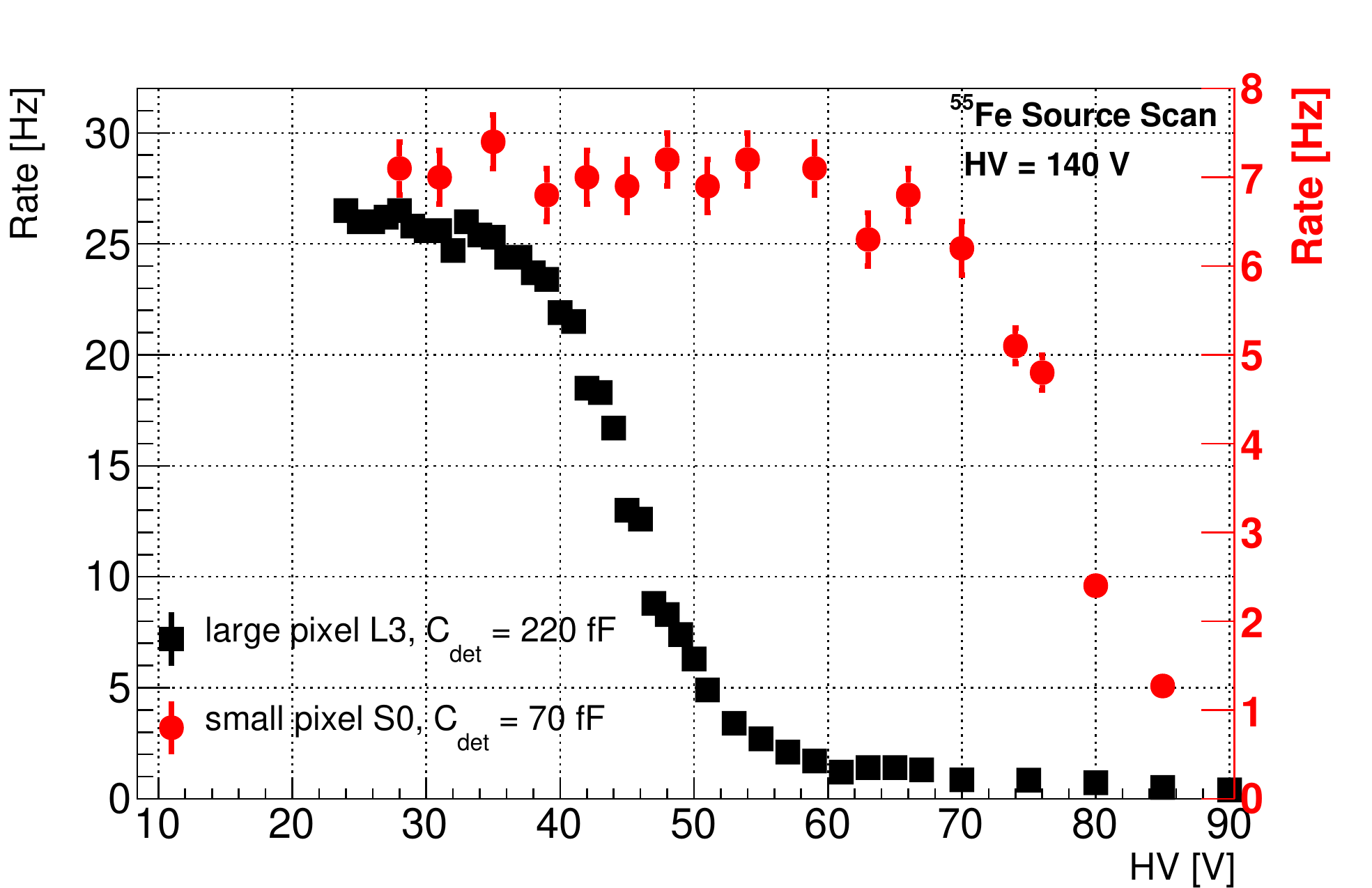}
\includegraphics[width=.49\textwidth,trim=0 0 0 0, clip]{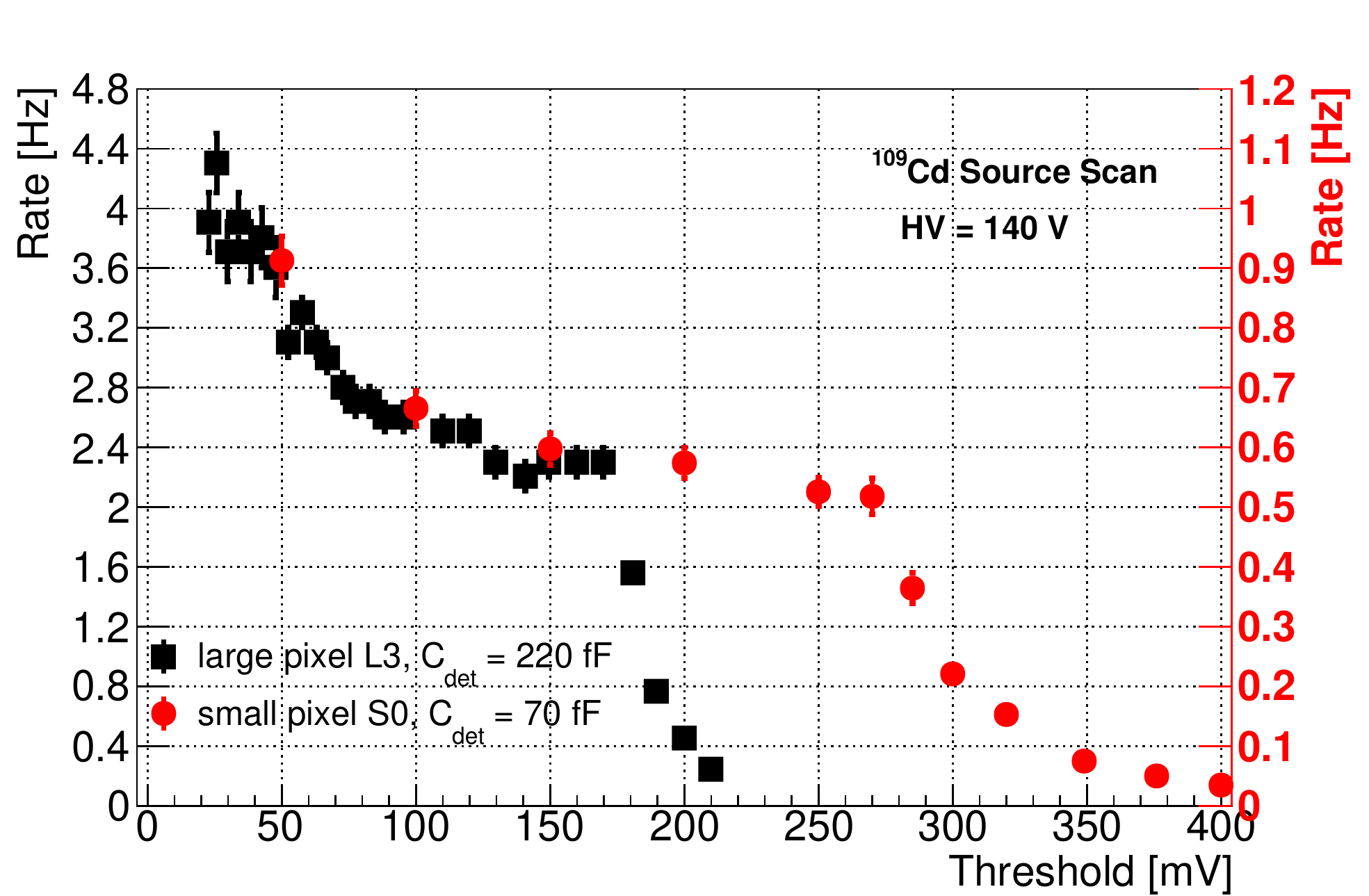}
\caption{\label{fig:rate} Photon hit rate vs. discrimination threshold as obtained with (left) a $ \mathrm{^{55}Fe} $ source and (right) a $ \mathrm{^{109}Cd} $ source for the small pixel S0 (red dots) and for the large pixel L3 (black squares) at HV = 140 V. 
The offset voltage of the amplifier output is subtracted to the threshold.
}
\end{figure}

Figure \ref{fig:rate} shows the photon hit rate as a function of the threshold measured with a $ \mathrm{^{55}Fe} $ and a $ \mathrm{^{109}Cd} $  source, for a high voltage of \SI{140}{\volt}. 
The rate is approximately constant for low threshold values, indicating a good discrimination of the photon peak. 
The photons from the $ \mathrm{^{109}Cd} $ source are energetic enough to allow  a rather accurate measurement  of the gain, which results to be 
$ A_{Q} = $ \SI{290}{\milli\volt\per\femto\coulomb} for the small pixel and \SI{185}{\milli\volt\per\femto\coulomb} for the large one.
Finally, from the measurement of the amplifier noise and the charge gain it is possible to estimate the ENC as: 
\begin{center}
$ ENC = \displaystyle\frac{\sigma_{V}}{A_Q} $
\end{center}
which gives an ENC of 90 electrons for the small pixel and 160 electrons for the large one. This results is ~30\% higher than the CADENCE simulation. 
This discrepancy, generated by the measured values of $A_Q$, 
could be explained either by a parasitic feedback capacitance of \SI{1.5}{\femto\farad} in the amplifier circuit or by a lower voltage gain of the amplifier.

\subsection{Time resolution}

The electrons emitted by the $ \mathrm{^{90} Sr} $ source were used to measure the TOF between pixel L3 and pixel S0 of the prototype and the reference LGAD detector. 
For this measurement the output of the pixel under test was read by an oscilloscope with 25 Gs/s sampling rate and analogue bandwidth limited to \SI{6}{\giga\hertz}. The remaining pixels were operated at low threshold and sent to a single channel of the oscilloscope to generate a tag signal for events with charge shared by the pixel under test and the neighbouring pixels.
In the case of the large-pixel matrix, pixel L5, which shares the same output of L3, was masked.

Data were taken at several threshold and HV values.      
%
The TOF values were calculated as: 
\begin{center}
$ \mathrm{TOF} = t_{\mathrm{hexa}} - t_{\mathrm{LGAD}} $ 
\end{center}
where $ t_{\mathrm{hexa}} $ is the time measured by the prototype chip and $ t_{\mathrm{LGAD}} $ is the time measured by the reference LGAD.
The selection cuts applied to the LGAD to obtain the sample used for the TOF were the same used for the verification of its 50 ps time resolution. 
No event selection was applied to the data of the prototype chip under test.

\begin{figure}[htbp]
\centering
\includegraphics[width=.47\textwidth,trim=0 0 120 50,clip]{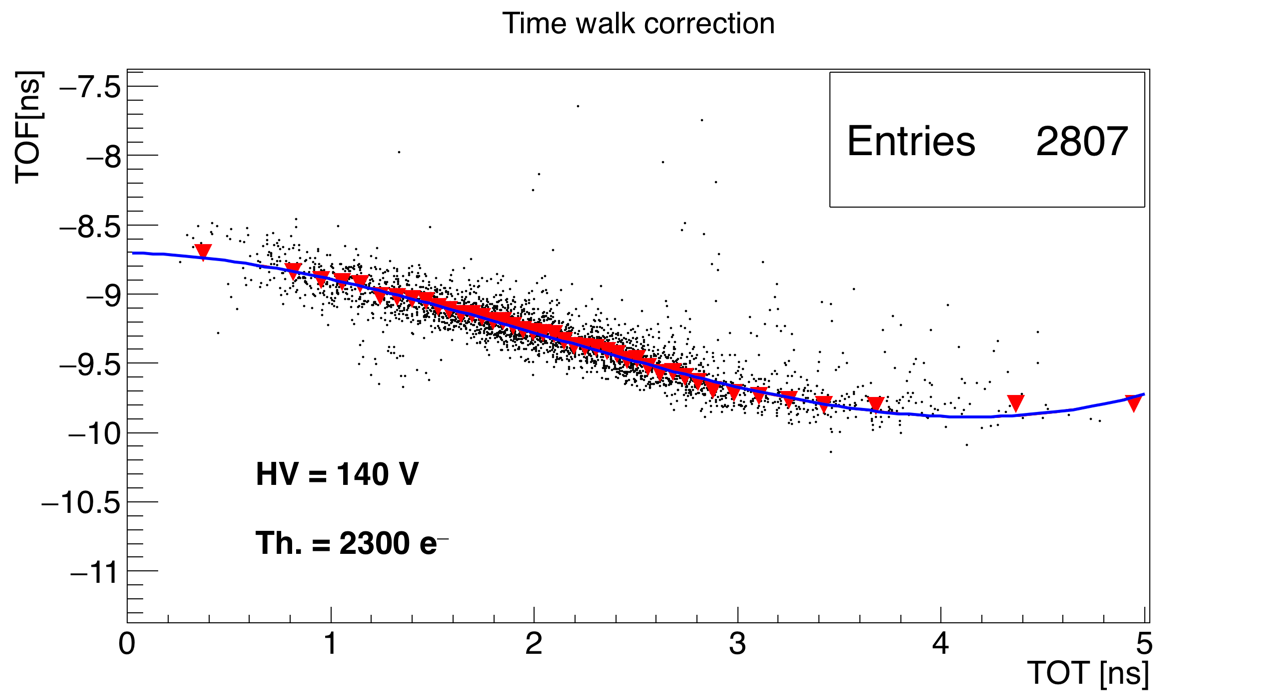}
\qquad
\includegraphics[width=.47\textwidth,trim=0 0 120 50,clip]{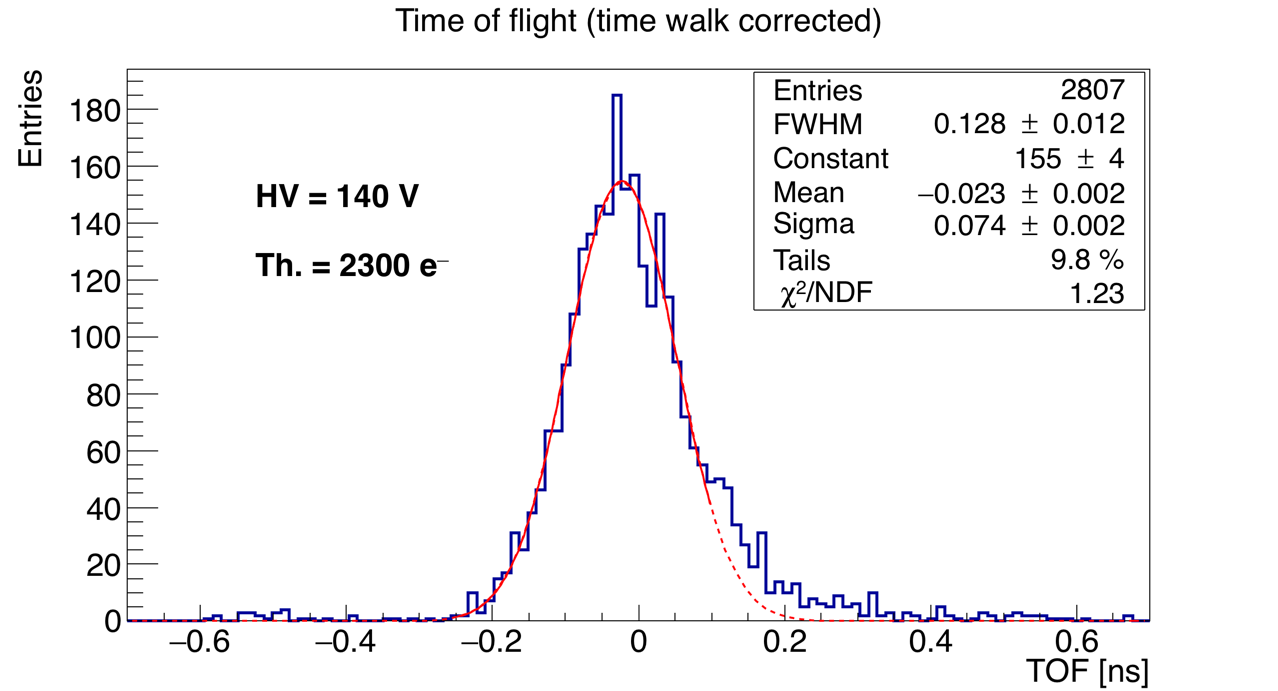}

\vspace{10pt}
\includegraphics[width=.47\textwidth,trim=0 0 120 50,clip]{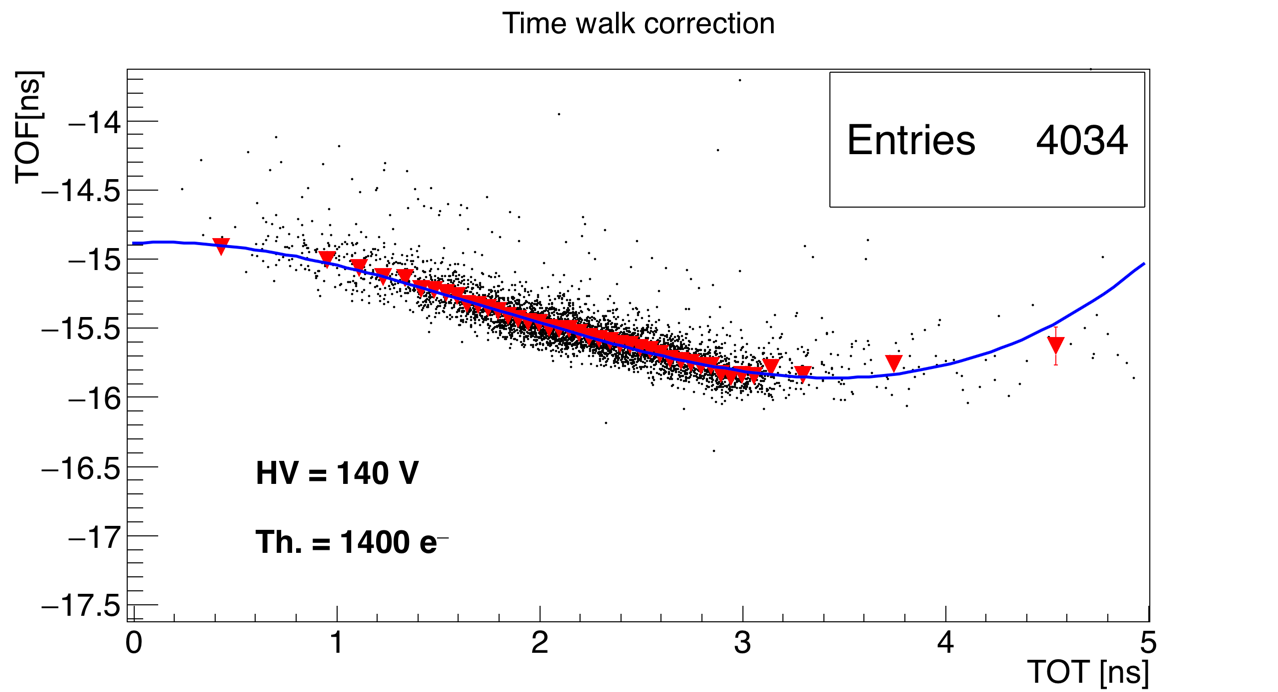}
\qquad
\includegraphics[width=.47\textwidth,trim=0 0 120 50,clip]{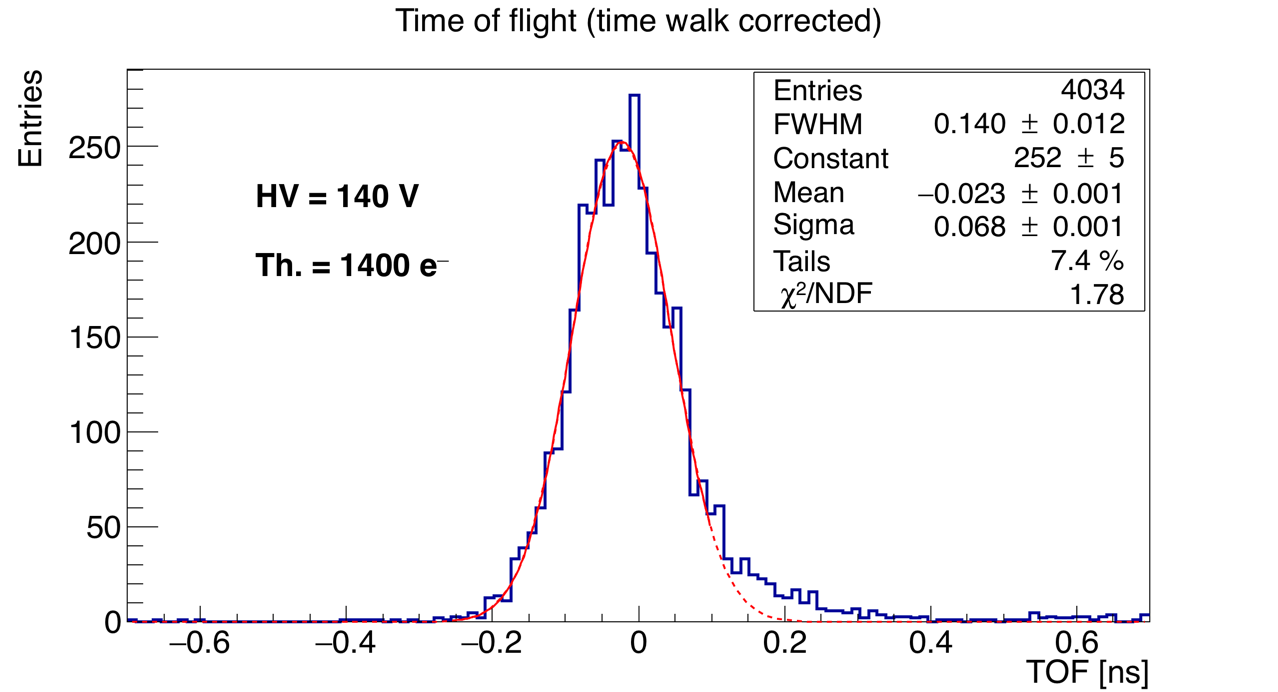}
\caption{\label{fig:TW} 
The two top panels show the data taken with the large hexagonal pixel L3 at HV = 140 V and threshold of 2300 electrons. The two bottom panels show the same information for the small pixel S0 at HV = 140 V and threshold of 1400 electrons. (Left) TOF  vs.\! the TOT  measured in the pixel.
The black dots represent the events, while the red triangles the average values of the TOF distribution in each bin of TOT. These averages  are used to compute the time-walk correction function shown by the blue line.
(Right) TOF distribution after time-walk correction.
The red line shows the results of the Gaussian fit performed using only the  interval TOF < +100 ps, marked by the full red line. The tail of the distributions outside the Gaussian fits amounts to 9.8 \% for the large pixel and to 7.4\% for the small pixel.
}
\end{figure}

The left panels of Figure \ref{fig:TW}  show the time-walk corrections obtained for the small pixel S0 and the large pixel L3 at  HV = 140 V and a threshold of 1400 and 2300 electrons, respectively.
The maximum time walk is found to be less than \SI{1}{\nano\second}. 

The right panels of Figure \ref{fig:TW}  show the TOF distribution for the same working points after time-walk correction.
A  non-Gaussian tail is observed for positive TOF values.
This tail, present for all threshold and HV values, could be attributed to the electrons from the source crossing the region between two pixels and requires to be investigated in a testbeam.
Gaussian fits to the time-walk corrected TOF distributions were performed
to establish the time resolution of the core of the distributions. 
TOF times larger than +100 ps were excluded from the fits to avoid that the asymmetric tail 
alters the results.
The time resolution of our  monolithic prototype chip was calculated by subtraction in quadrature of the 50 ps time resolution contribution of the reference LGAD from the time-walk corrected TOF value. 
For the small pixel at the working point of Figure~\ref{fig:TW} bottom we obtain
\begin{center}
$\sigma = \sqrt{\sigma_{\mathrm{TOF}}^2 - \sigma_{\mathrm{LGAD}}^2} = \sqrt{68^2-50^2} = (46 \pm 1 )$ ps
\end{center}
while for the large pixel at the working point of Figure~\ref{fig:TW} top the Gaussian part of the time resolution provides $\sqrt{74^2-50^2} = (55 \pm 2)$ ps.
 In all instances 
the non-Gaussian tails were below 10 \%. 
 The negative mean value of the time-walk corrected TOF distributions is an artefact  due to the fact that the time-walk corrections were calculated using all the events, including those in the tails.

\begin{figure}[htbp]
\centering 
\qquad
\includegraphics[width=.74\textwidth,trim=0 0 0 20, clip]{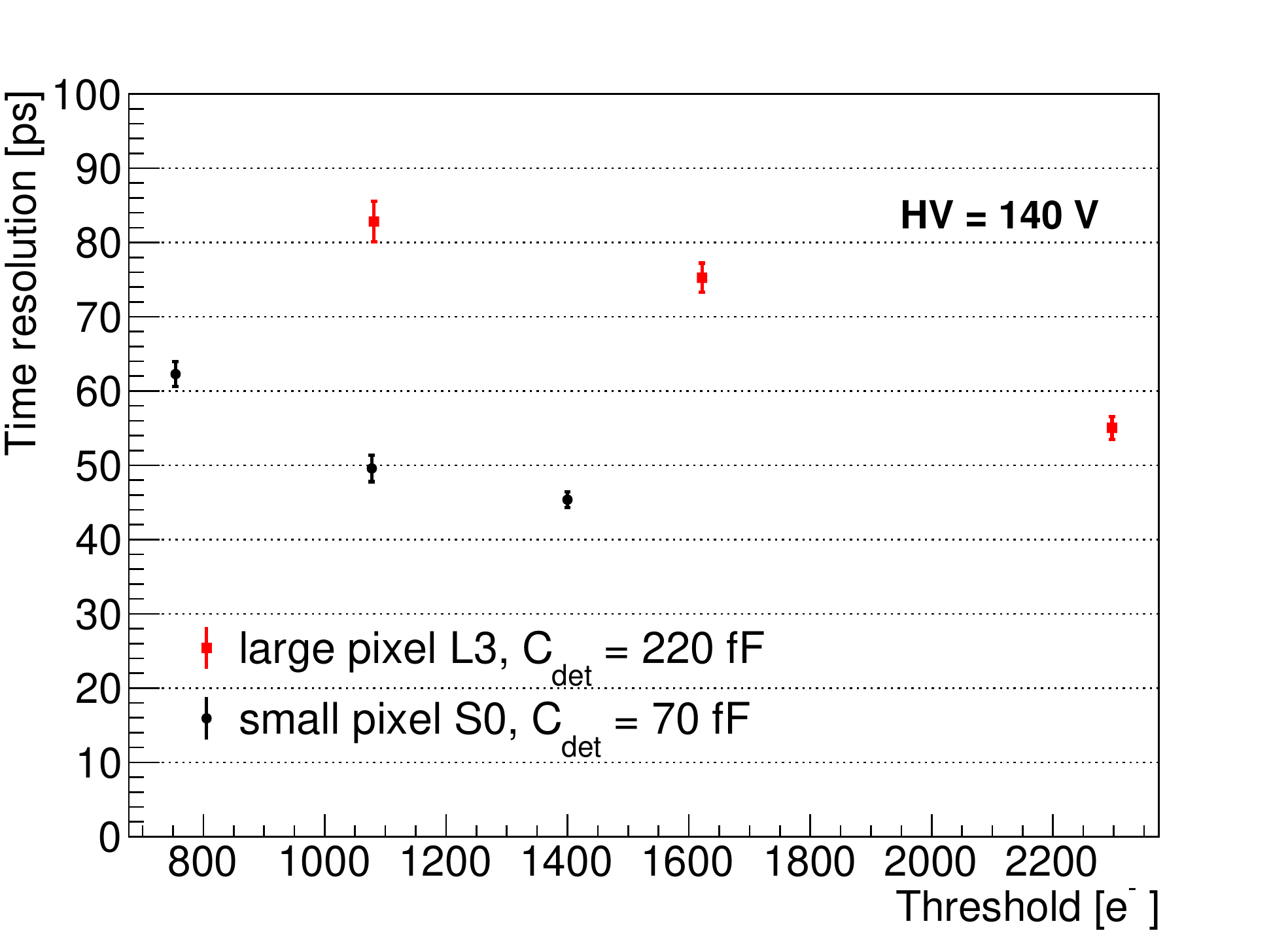}
\caption{\label{fig:TOF_SCAN} Time resolution of the large pixel L3 (red squares) and the small pixel S0 (black dots) of the prototype monolithic detector as a function of the discrimination threshold at HV = 140 V.}
\end{figure}

Figure \ref{fig:TOF_SCAN} shows the time resolution of the prototype as a function of the discrimination threshold for the small pixel S0 and the large pixel L3 operated at HV = 140 V. 
For both pixel sizes  the time resolution is measured to improve with increasing threshold.
This observation could be explained by an improved resolution in the time-walk correction at high threshold values: the TOT, used to correct for the time walk, has a smaller error when the analogue signal crosses  the discrimination threshold with a larger slope.  
Therefore, the present timing performance appears to be limited by the method selected to correct for the time walk.

Although the detection efficiency 
cannot be measured with 
the present experimental setup using a radioactive source,
it is possible to 
verify that the threshold value is not cutting the measured-charge distribution above its most-probable  value.
The depletion depth of a silicon sensor with bulk resistivity of \SI{50}{\ohm\centi\meter} operated at 140 V is \SI{26}{\micro\meter}, which corresponds to a most probable deposited charge from a MIP of approximately 1600 electrons.
Therefore, we should expect that in Figure \ref{fig:TOF_SCAN} only the two higher-threshold measurements for the large pixel suffer from a significant loss in efficiency in the detection of MIPs.
For the small pixel, at the intermediate threshold value, which should be close to maximal efficiency operation, the measured time resolution is $(50\pm1)$ ps. 
For a similar threshold the large pixel shows a time resolution of $(83\pm3)$ ps.

\begin{figure}[htbp]
\centering 
\qquad
\includegraphics[width=.74\textwidth,trim=0 0 0 20, clip]{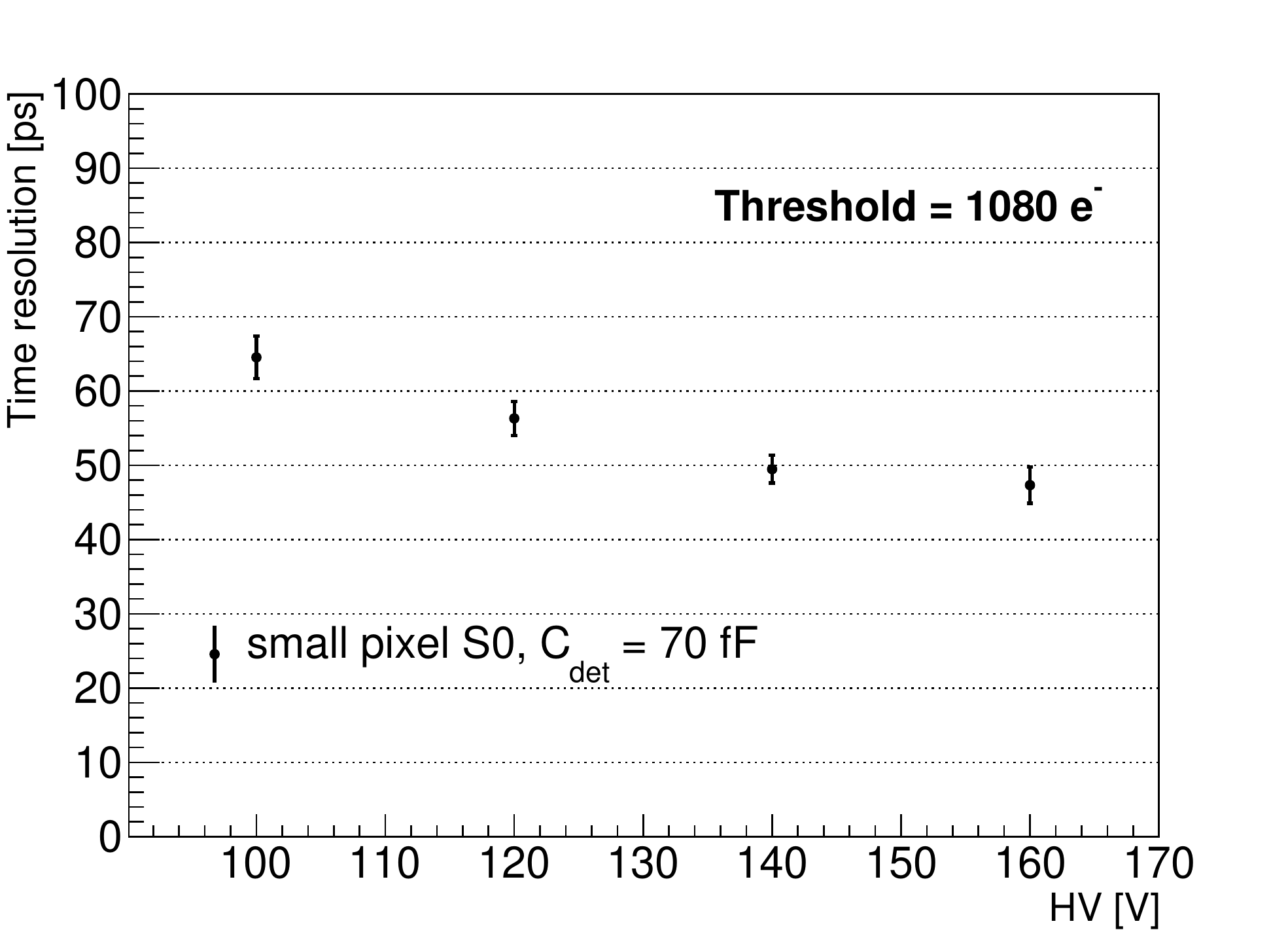}
\caption{\label{fig:TOF_SCAN_HV} Time resolution of the small pixel S0  as a function of the high voltage for a discriminator threshold of 1080 electrons.}
\end{figure}

Figure \ref{fig:TOF_SCAN_HV} shows the dependence of the time resolution on the high voltage for the small pixel.
An improvement of the time resolution is observed with increasing high voltage.
This behaviour could be attributed to: 
{\it i)} an increase of the depletion region that provides a larger charge and a slightly lower capacitance, and therefore larger signal/noise; {\it ii)}  a better uniformity of the carrier drift velocity.
These limiting factors can be controlled by producing the sensor on a high resistivity epitaxial layer with a highly doped substrate.

\section{Conclusions}
\label{sec: conclusions}

A  monolithic pixel detector prototype was produced in the IHP SiGe BiCMOS  SG13G2 technology, with a sensor design optimised for timing and a fast and low-noise SiGe HBT  amplifier. 
The prototype contains small matrices of hexagonal pixels  of \SI{130} and  \SI{65}{\micro\meter} side and  capacitance of   approximately \SI{220} and \SI{70}{\femto\farad}, respectively.
Single-hit rate scans performed  with a  $ \mathrm{^{109}Cd} $ source indicate an ENC of the preamplifier of 90 electrons RMS and a gain of \SI{290}{\milli\volt\per\femto\coulomb}  for the small pixels, and 160 electrons RMS and   \SI{185}{\milli\volt\per\femto\coulomb}  for the large pixels.

Data  taken with a 
$ \mathrm{^{90} Sr} $ source were used to measure the TOF between the prototype and a reference LGAD detector that has 50 ps time resolution. 
A total excursion of the time walk below 1 ns was measured, which confirms the  fast response of the amplifier.
At a bias voltage of 140 V, time resolutions between $(62\pm2)$ ps and $(46\pm1)$ ps were measured at different thresholds for the smaller pixels after time-walk correction. These resolutions refer to the $\sim$90\% of the events in the Gaussian core of the distributions. No event selection was performed.
These results are competitive with silicon technologies that integrate an avalanche mechanism,
and prove that SiGe HBT technology can provide precision tracking and excellent time resolution even in the absence of an internal gain mechanism.
Experiments at beamline facilities with an external tracking system will allow for more detailed studies and for an accurate measurement of the efficiency.



\acknowledgments

The authors wish to thank  Ivan Peric for several useful discussions on the sensor integration in the CMOS process, and Nicol\`o Cartiglia for providing the reference LGADs used in this measurement. Special thanks go to the technical staff of the DPNC of the University of Geneva for the assembly of the instrumentation and for the support during the preparation of the test.
This research is based on the work performed within the TT-PET project funded by the Swiss National Science Foundation grant CRSII2-160808.



\end{document}